\documentclass[twocolumn,groupedaddress]{revtex4}
\usepackage{graphicx}

\begin{document}

\title{Reading and writing charge on graphene devices}
\author{M. R. Connolly$^{1,2}$, E. D. Herbschleb$^1$, R. K. Puddy$^1$, M. Roy$^3$, D. Anderson$^1$, G. A. C. Jones$^1$, P. Maksym$^3$, C. G. Smith$^1$}
\affiliation{$^1$Cavendish Laboratory, Department of Physics, University of Cambridge, Cambridge, CB3 0HE, UK}
\affiliation{$^2$National Physical Laboratory, Hampton Road, Teddington TW11 0LW, UK}
\affiliation{$^3$Department of Physics and Astronomy, University of Leicester, University Road, Leicester, LE1 7RH, UK}

\date{\today}

\begin{abstract}
We use a combination of charge writing and scanning gate microscopy to map and modify the local charge neutrality point of graphene field-effect devices. We give a demonstration of the technique by writing remote charge in a thin dielectric layer over the graphene-metal interface and detecting the resulting shift in local charge neutrality point. We perform electrostatic simulations to characterize the gating effect of a realistic scanning probe tip on a graphene bilayer and find a good agreement with the experimental results.
\end{abstract}


\maketitle

Gate tuneable conductivity underpins the operation of a multitude of applications envisaged for graphene, ranging from switching in mass-produced field-effect transistors to single electron manipulation in quantum dots. Topographic, morphological, and electrostatic variability often compromises device performance, so developing and adapting techniques for imaging and modifying local electronic properties continues to receive considerable interest. One such technique is charge writing (CW), which uses the biased tip of a scanning probe microscope to deposit charge in a dielectric layer electrostatically coupled to a nanodevice, mainly those fabricated from two-dimensional electron gas-based systems in semiconductor heterostructures \cite{Crook2003,Crook2006}. Scanning gate microscopy (SGM) employs the same tip as a mobile gate for imaging the charge-induced changes in local conductivity. While SGM has also recently proven capable of making charge neutrality point (CNP) maps of inhomogeneously doped graphene flakes \cite{Connolly2010,Chen2010}, CW has yet to be used to modify the local CNP. In this letter we measure and modify the local CNP using combined CW/SGM, paving the way towards \textit{in situ} electrostatic patterning of arbitrary potential landscapes in graphene field-effect devices.

\begin{figure}
\includegraphics{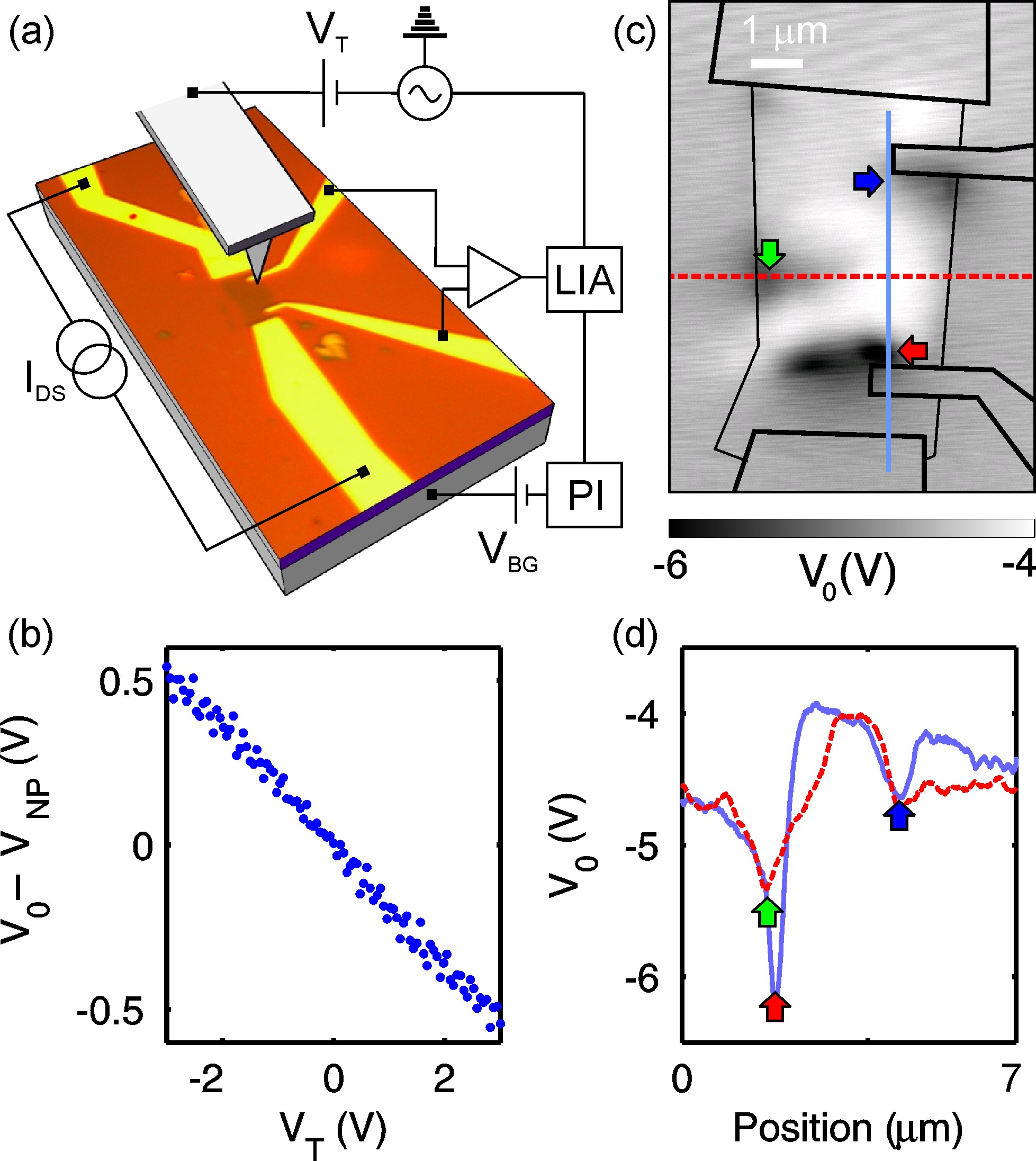}
\caption{(a) Circuit used to perform nulling scanning gate microscopy in four-terminal configuration on a graphene bilayer [LIA: Lock-In Amplifier, PI: Proportional-Integral controller.] (b) Difference between the nulling back-gate voltage and the bulk neutrality point voltage as a function of d.c. voltage applied to the tip. (c) Map of the neutrality point $V_0$ of the bilayer. (d) Horizontal (dashed) and vertical (solid) line profiles. Arrows correspond to the $n$-type doping features indicated in the map.}    
\label{Fig:Fig1}
\end{figure}

Our graphene devices are fabricated from flakes which are mechanically exfoliated from natural graphite onto a highly doped Si substrate capped with a 300 nm thick SiO$_{2}$ layer. The number of layers is identified from their optical contrast and 5/30 nm Ti/Au contacts are patterned using e-beam lithography, thermal evaporation, and standard PMMA lift-off processing. For improved stability and sensitivity we operate our scanning probe microscope in non-contact mode (see \cite{Connolly2010} for details) and under vacuum ($10^{-5}$ mbar) with an average tip-surface seperation of 5-20 nm. The SGM setup used to image the CNP is shown in Fig. \ref{Fig:Fig1}(a). To benefit from the high signal-to-noise ratio achievable using a.c. detection, we modulate the voltage V$_T$ on the tip (NanoWorld ARROW-NCPt) at low frequency (typically 3 V @ 1 kHz) and detect the modulation of $I_{DS}$ ($\approx$ 250 $\mu$A) using a lock-in amplifier. The demodulated component $\delta I_{DS}$ is proportional to the local transconductance $ g_m =\partial I_{DS}/\partial V_T$ ($\approx$ 0.2 $\mu$A/V) \cite{Crook2000, Connolly2010}. Our earlier work established that the back-gate voltage which nulls the transconductance is the local charge neutrality point ($V_0$), and spatial maps of $V_0$ were generated by performing point spectroscopy and extracting $V_0$ at each point \cite{Connolly2010}. Owing to the sign change of the transconductance either side of $V_0$, $g_m(V_{BG})$ is highly suited as an error signal which can be nulled in a feedback loop for real-time tracking of $V_0$, similar to the compensation of the contact potential in scanning Kelvin probe microscopy \cite{Nonnenmacher1991}. Fig. \ref{Fig:Fig1}(a) shows the circuit used to implement ``nulling'' scanning gate microscopy (NSGM). The modulation in current $\delta I_{DS}$ is fed into a software controlled feedback loop which adjusts the back-gate voltage to maintain a constant setpoint $\delta I_{DS}(V_0)$=0. Fig. \ref{Fig:Fig1}(c) shows a map of $V_0$ captured using NSGM in four-terminal configuration on a bilayer flake, revealing up to 2 V variations in $V_0$. Most notably, regions with pronounced $n$-type doping exist around the Ti/Au contacts. Line profiles shown in Fig. \ref{Fig:Fig1}(d) show that the doping extends hundreds of nanometers away from the contacts. We have performed NSGM on other flakes in both two and four terminal configurations and observe similar doping profiles around contacts, in good agreement with various microscopy studies \cite{Lee2008, Mueller2009, Yu2009} and theoretical analyses taking into account the different work functions of graphene and titanium \cite{Khomyakov2010}. 

\begin{figure}
\includegraphics{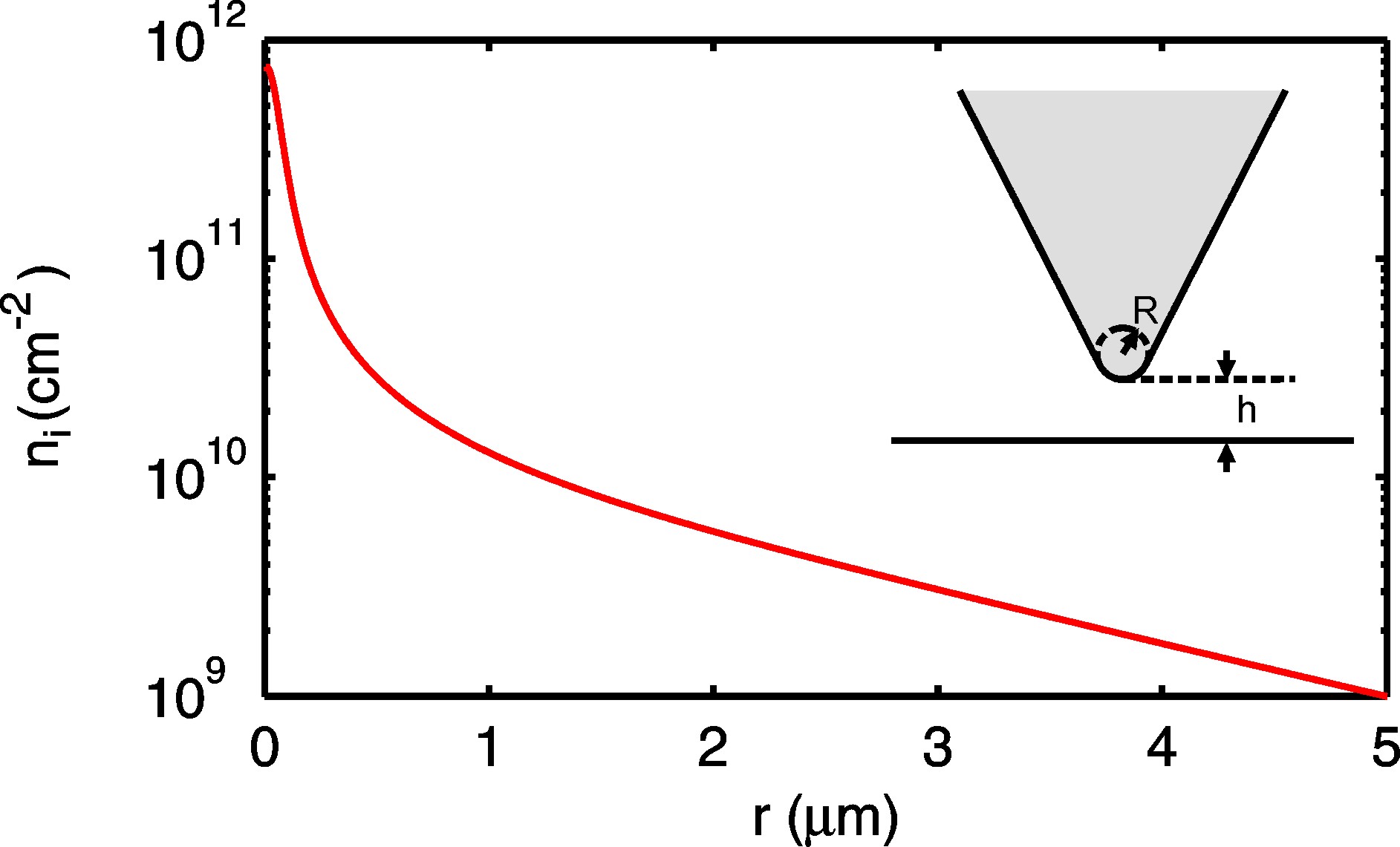}
\caption{Main figure: calculated induced electron density. Inset: schematic of tip
geometry ($h=20$ nm). The tip is modelled as a blunt cone with an opening angle 
of 30${^\circ}$ and a spherical end with $R=100$ nm as shown. }    
\label{Fig:Fig2}
\end{figure}

The features in the NSGM images shown in Fig. \ref{Fig:Fig1}(b) vary over hundreds of nanometers, indicating a large effect from the conical section of the tip. This is not unexpected as the tip used in these experiments has a relatively large half-cone angle $\theta_0 \approx$ 30$^\circ$, which starts to dominate for tip-surface separations $\geq$5 nm \cite{Wilson2008, Hudlet1998}. A similar long-range SGM response was reported in the context of SGM on carbon nanotubes \cite{Wilson2008} and subsurface two dimensional electron gases \cite{Baumgartner2006, Ouisse2008}, where it was distinguished from finer conductance variations which occur on the scale of the tip apex \cite{Ouisse2008}. 

In addition to the length scale of the perturbation, an important parameter characterizing SGM is the carrier density induced under the tip, which can be quantified through the effective capacitive coupling $\beta=\Delta n/V_T$, where $\Delta n$ is the change in carrier density under the tip. NSGM enables us to measure $\beta$ directly by sweeping the DC voltage applied to the tip and tracking the nulling back-gate voltage. Fig. \ref{Fig:Fig1}(b) shows a plot of the difference between the nulling voltage and the bulk neutrality point voltage ($V_{NP}$) as a function of $V_T$. Using the known relationship between carrier density and back-gate voltage, $n=\alpha (V_{BG}-V_{NP})$ ($\alpha$ = 7.2 $\times$ 10$^{10}$ cm$^{-2}$) \cite{Novoselov2004}, we infer a coupling between the tip and the graphene of $\beta$$\approx$ 13 $\times$ 10$^{9}$ cm$^{-2}$/V. 

To obtain a theoretical estimate for the length scale and coupling strength of the tip perturbation, we solve Laplace's equation numerically in cylindrical polar co-ordinates for a realistic tip-sample geometry. We include the screening charge induced in the graphene self-consistently within the Thomas-Fermi approximation. In our model, a blunt tip (see inset to Fig. \ref{Fig:Fig2}) is positioned 20 nm above a clean, homogeneous bilayer graphene sheet which lies on a 300 nm thick dielectric substrate. Experimentally we are close to the neutrality point so, in our idealised model, we fix the voltage at the bottom of the substrate to 0 V. At the upper boundary of the dielectric, the electric displacement is discontinuous by an amount determined by the charge density induced in the bilayer. We calculate this self-consistently. The Fermi level, $E_{F}(r)$, in the bilayer varies spatially with the voltage in the graphene and the induced charge density, $n_i (r) = \gamma_1 (E_F(r)-E_g(r))/(2\pi P^2)$ for $E_F > E_g/2$, is derived from the low energy parabolic dispersion for a bilayer \cite{McCann2006-1}. Here, $P=0.539$ eV nm \cite{DiVincenzo1984}, $\gamma_1 = 0.39$ eV \cite{McCann2006-1} and $E_g(r) = [e^2V_{bl}^2 \gamma_1^2 / (e^2V_{bl}^2+\gamma_1^2)]^{1/2}$ \cite{Castro2007} is the band gap induced in the bilayer due to the bias, $V_{bl}$, across it. 

The calculated charge density is shown in Fig. \ref{Fig:Fig2}. In our calculation we extend the system in radial (to $r=50$ $\mu$m) and vertical (to $z=6.3$ $\mu$m) directions until $n_i(r)$ is converged to within $0.05\%$ for $r<5$ $\mu$m. Qualitatively, $n_i(r)$ is not sensitive to the band-structure model. Even changing the bilayer to monolayer graphene 
has an effect of less than a few percent. Instead, it is the tip shape and $V_T$ that primarily determine $n_i(r)$. Within $r<5$ $\mu$m we find that the charge density is well fitted by the sum of two Lorentzians. The Lorentzian peak height varies linearly with $V_T$, but the half widths are insensitive to tip voltage and vary by $<\! 1\%$ for $0.5\! \leq\!  V_T\! \leq\! 3$ V. At $V_T=3$ V the wider of the Lorentzians has a peak height, $n_i(0) = 14\times10^{9}$ cm$^{-2}$, and a half width of 3.6 $\mu$m. This fits well with the observed micron-sized perturbation and gives a value for $\beta= \Delta n/V_T =n_i(0)/V_T\approx 5 \times 10^{9}$ cm$^{-2}$/V, similar to that found experimentally.

\begin{figure}[!h]
\includegraphics{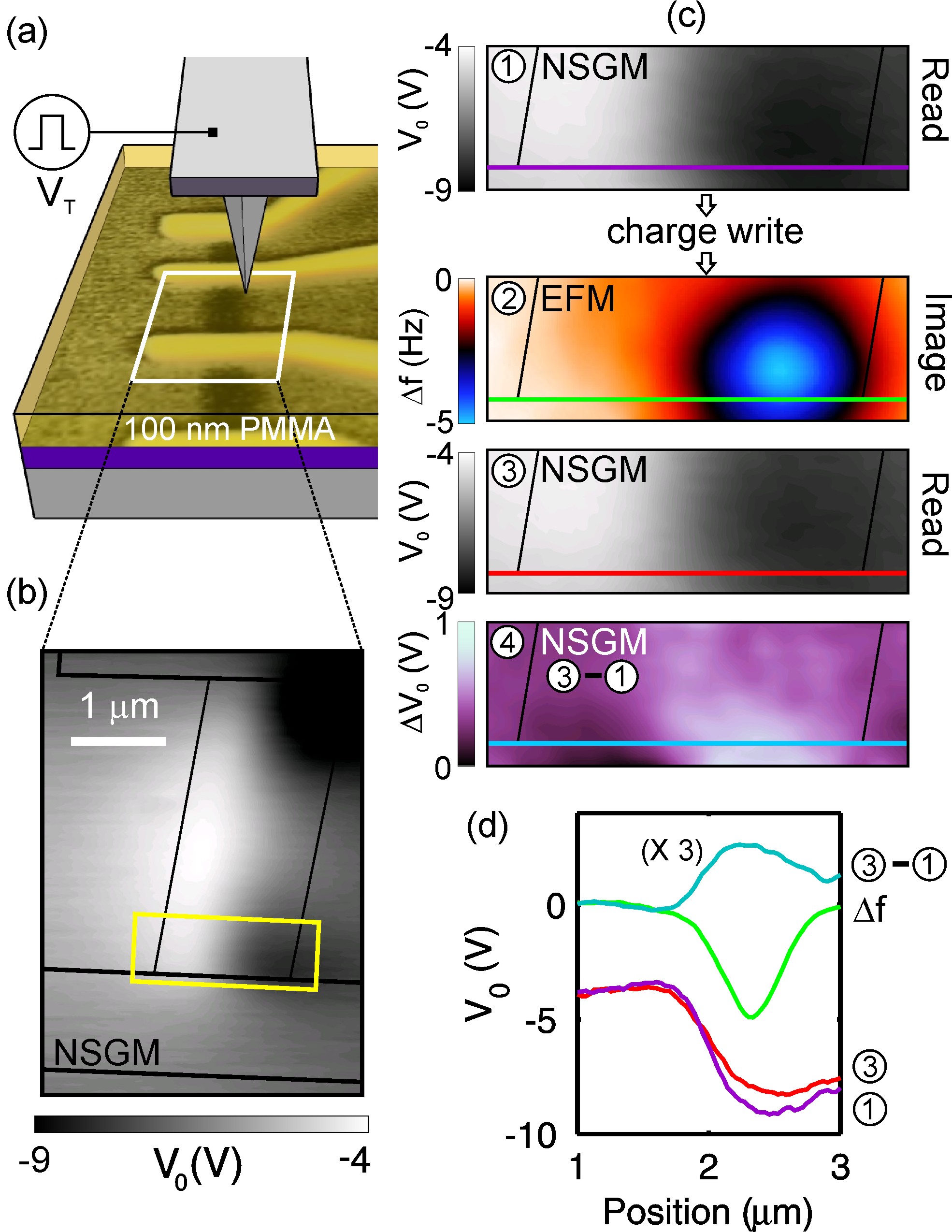}
\caption{(a) Setup used to write charge in a 100 nm thick layer of PMMA spin-coated over the graphene. Voltage pulses (20 V$_{Pk-Pk}$ @ 50 Hz) are applied to the tip in tapping mode to write charge. (b) Nulling scanning gate micrograph showing the charge neutrality point distribution prior to writing. (c) Procedure for combining NSGM with charge writing to sequentially read and write charge neutrality point distributions over graphene devices. (d) Line profiles of the neutrality point and frequency shift for the NSGM and EFM images, respectively.}    
\label{Fig:Fig3}
\end{figure}

We now show how the tip can be employed to modify the CNP map by performing CW to induce holes in the graphene and thereby partially compensate for the $n$-type doping around the contacts. We spin coated a 100 nm thick layer of poly(methyl methacrylate) (PMMA) to act as the CW dielectric over the device shown in Fig. \ref{Fig:Fig3}(a). Fig. \ref{Fig:Fig3}(b) shows the NSGM map in the presence of the PMMA layer. Note the reduced spatial resolution due to the necessity of scanning $\approx$ 100 nm from the surface, and the enhanced $n$-type doping, especially next to the contacts where $V_{0}$ reaches as low as $\approx$ -9 V. Further work is required to use thinner and less invasive dielectrics. Prior to CW we acquire a high resolution NSGM image of $V_{0}(x,y)$ next to the contacts [Fig. \ref{Fig:Fig3}(c), image 1]. To perform CW in PMMA we follow the procedure presented in \cite{Ressier2008} for creating electrostatic templates. The tip is brought in to hard tapping-mode contact over the doped region and a square-wave signal between -10 and -30 V at 50 Hz is applied to the tip. These values bring the pulses over the $\approx$ 15 V threshold pulse amplitude for charge deposition in PMMA \cite{Ressier2008}. As expected in tapping mode and pulsed writing, topographic imaging did not show any subsequent damage to the PMMA \cite{Ressier2008}. For imaging the deposited charge we employ electrostatic force microscopy (EFM), and Kelvin probe microscopy when quantitative estimates of the surface potential are required. An EFM image showing the shift in frequency ($\Delta f$) of the cantilever over the charged region is shown in Fig. \ref{Fig:Fig3}(c), image 2. The charge is well localized and corresponds to a surface potential of $\approx$ -2 V. We re-map $V_0$ in NSGM after CW (Fig. \ref{Fig:Fig3}(c), image 3) and construct the difference (Fig. \ref{Fig:Fig3}(c), image 4) to enhance the CNP shift. Line profiles in Fig. \ref{Fig:Fig3}(d) show a good correspondence between the peaks of written charge and the CNP shift, which is $\approx$ 800 mV. We found the charge to be stable for several days under vacuum at room temperature, in good agreement with previous observations \cite{Ressier2008}.   

In conclusion, we have introduced a method combining charge writing and scanning gate microscopy to read and write the local charge neutrality point on a graphene device. We mapped the variation of doping across a bilayer flake and found micron-sized $n$-type doping from Ti/Au contacts. We performed electrostatic simulations to analyze the interaction between the biased tip and the graphene and deduced a capacitive coupling of $\beta$$\approx$ 5$\times$10$^{9}$ cm$^{-2}$/V which is in reasonable quantitative agreement with the value determined experimentally. We deposited charge with a surface potential of $\approx$ 2 V and shifted the CNP of a region next to the metallic contacts by $\approx$ 800 mV. The use of thinner, less invasive dielectrics with higher permittivity would enable the \textit{in situ} fabrication of graphene devices with arbitrary potential landscapes, and band-gaps in the case of graphene bilayers \cite{Szafranek2010}.

This work was financially supported by the European GRAND project (ICT/FET) and EPSRC. 


\end{document}